\documentclass[%
 aip,
 amsmath,amssymb,
 reprint,%
]{revtex4-1}

\usepackage{graphicx}
\usepackage{dcolumn}
\usepackage{bm}
\usepackage[utf8]{inputenc}
\usepackage[T1]{fontenc}
\usepackage{mathptmx}
\usepackage{etoolbox}
\usepackage{adjustbox}
\usepackage{csquotes}
\makeatletter
\def\@email#1#2{%
 \endgroup
 \patchcmd{\titleblock@produce}
  {\frontmatter@RRAPformat}
  {\frontmatter@RRAPformat{\produce@RRAP{*#1\href{mailto:#2}{#2}}}\frontmatter@RRAPformat}
  {}{}
}%
\makeatother
\begin{document}
	\title{ Electronic transport in copper-graphene composites  }  
	\author{Kashi N. Subedi}

	\affiliation{ 
		Department of Physics and Astronomy, Ohio University,  Athens, OH 45701, USA 
	}%
	\author{Kishor Nepal}%
	\affiliation{ 
	Department of Physics and Astronomy, Ohio University,  Athens, OH 45701, USA 
}%
	\author{Chinonso Ugwumadu}%
	\affiliation{ 
	Department of Physics and Astronomy, Ohio University,  Athens, OH 45701, USA 
}%
    	\author{Keerti  Kappagantula}
    \affiliation{ 
    	Pacific Northwest National Laboratory, Richland, WA 99352, USA 
    }%
	\author{D. A. Drabold}
	\altaffiliation[Email :  ]{drabold@ohio.edu}
	\affiliation{ 
		Department of Physics and Astronomy, Ohio University,  Athens, OH 45701, USA 
	}%

	\date{\today}

\begin{abstract}
We investigate the electronic transport properties of copper-graphene composites using a density-functional framework.~Conduction in composites by varying the interface distance of a copper/graphene/copper (Cu/G/Cu) interface models was studied.~The electronic density of states reveals increasing contributions from both copper and carbon atoms near the Fermi level with decreasing Cu-G interfacial distance.~Electronic conductivity of the models  computed using the Kubo-Greenwood formula  showed the conductivity increases with decreasing Cu-G distance.~We also find that the conductivity saturates below a threshold Cu-G distance.~By computing the space-projected conductivity of the Cu/G/Cu models, we show that the graphene forms a bridge to the electronic conduction at small copper-graphene distances, thereby enhancing the conductivity.
\end{abstract}

\maketitle

~Metal composites are key industrial materials that may be manufactured using various novel techniques~\cite{HERINGHAUS_1996,Bouaziz_2013,Bruschi_2021}, leading to metastable materials exhibiting desirable mechanical, thermal, electronic, and transport properties without the constraints typically seen in alloys.~Copper is used extensively for energy transport and electromagnetic applications such as motors, generators, cable, busbars and transformers.~In the last two decades, we have witnessed copper composites with improved transport properties~\cite{Manvandra_2018}, devised as a means to reduce energy loss in applications.\\

One material that has demonstrated improved electrical conductivity and current density is a copper-graphene (Cu-G) composite. Various Cu-G composites ranging from a simple interface laminated to bulk Cu with interconnected graphene flakes have been reported in the literature, with some showing enhanced conductivity~\cite{Weiping_2015,Hokyun_2017,Zhonglie_2019,Yang_2021}, and more commonly enhancement in mechanical performance~\cite{Hwang_2013,Zhonglie_2019}.~Conventional wisdom suggests that the introduction of additives to metal would increase scattering that would lead to increased electrical resistivity~\cite{Ellis_1993}. This is the reason why addition of silver to copper leads to reduced electrical conductivity of Cu, even though Ag is 8\% higher in conductivity than Cu~\cite{Felicia_2018}.~The additives, on the other hand, also arrest dislocation movement in the metals due to several mechanisms such as precipitate hardening, solid solution strengthening and dispersion strengthening, resulting in enhanced mechanical performance~\cite{Kuhlmann_1989}.\\

At this juncture, it is known that the electrical conductivity of the Cu-G composites are dependent on the quantity of graphene used in the composites, the type of graphene and their defect density, and the arrangement of Cu matrix and graphene in the composite microstructure, which eventually dictates the carrier transport in the material during conduction~\cite{Keerti_2022, Pan_2022}.~Processing routes adopted to manufacture the composites are also known to influence electrical performance.~Processes that can control atomic scale deposition of Cu and graphene, such as vapor deposition and molecular beam epitaxy, have shown promise in manufacturing nano-to-micron scale samples demonstrating enhanced electrical performance~\cite{Zheng_2018}.~At the bulk scale, solid phase processing techniques have shown promise in enhancing electrical conductivity~\cite{Keerti_2022}.\\

In all this, however, while there is an evolving understanding of processing approaches and material chemistries that result in Cu-G composites with improved electrical performance, little is known about the nature of charge-carrier transfer between Cu and graphene.~Several papers discuss carrier mobility in graphene that is deposited on Cu foils~\cite{ Orofeo_2012,Banszerus_2015}.~However, limited experimental and computational research is published on transport phenomena across metal/graphene interfaces.~There is minimal understanding on how the nature of the interface (such as atom arrangement, interfacial distance) affects the electron density at the interface.~In Cu-G composites demonstrating enhanced conductivity compared to the corresponding Cu substrates, two causes were hypothesized for the enhancement in conductivity. In one case, where the manufacturing conditions allow for it, the Cu  grains are surmised to be templated on the adjacent graphene flakes leading to the formation of predominantly 111 textures in the composite during the manufacturing process. Such crystallographic orientations in Cu show higher conductivity compared to other textures such as 200 or 110. If there is a large concentration of 111 grains in the Cu microstructure, it is reasonable to expect an enhanced conductivity overall. In this hypothesis, graphene is thought to only template the grains but not actually participate in the conduction process. In the second case, graphene is assumed to participate during conduction through the exchange of carriers with the surrounding Cu matrix. One key consideration for this hypothesis is the nature of interface between the Cu and graphene that may be suitable for carrier transport across the interface. Some papers indicate that the 111 Cu grains offer the least lattice mismatch with graphene, which may also help with carrier transport. However, it is currently not well known how much conductivity the introduction of graphene brings about for Cu per this hypothesis. \\

In this Letter, we explore the electrical transport properties of the Cu-G composites as a function of the interfacial distance between Cu and graphene, a parameter that can be modulated during Cu-G manufacturing and a simple proxy for a possible strained local configuration.~We quantify the electronic transport by computing the electronic conductivity using the Kubo–Greenwood formula (KGF)~\cite{Kubo_1957,Greenwood_1958}, ideally suited for density-functional simulations of materials with its single particle Kohn–Sham orbitals and energies~\cite{Martin_2008}.~The KGF method has been utilized in tight-binding and DFT computations of the conductivity of liquids and solids~\cite{Allen_1987, Galli_1989}.~Besides quantifying the conductivity, we determine the conduction-active sites in the Cu-G composites by projecting the KGF conductivity onto real space using the space-projected conductivity (SPC), described elsewhere~\cite{Prasai_2018}.~The method has been implemented to study conduction processes in crystalline systems with defects, amorphous and semi-conducting systems at atomistic level~\cite{Subedi_2022,Thapa_2022,Subedi_2019,subedi_pssb}. \\

We simulated the Cu-G composite by constructing an interface model with a geometry of the form copper/graphene/copper (Cu/G/Cu) as shown in Figure 1a. We adopted one of the low-indexed surfaces, namely 111, to construct the Cu surface.~We placed the graphene sheet with one sub-lattice site (A) above the top of the first layer of Cu 111 and the other sub-lattice site (B) placed above the third layer of Cu 111 as shown in Figure~1b. This arrangement is also known as top-fcc, and has shown to represent the low-energy structure for Cu-G surface models~\cite{yang_2017_ab_initio}.~We simulated the external pressure on Cu-G composites by varying the copper-graphene distance ($d_{Cu-G}$) of the Cu/G/Cu interface model as shown in Figure 1a.~Besides Cu/G/Cu models, we also constructed an orthorhombic supercell of 216 atoms and simulated the external pressure by reducing the vertical dimension of the supercell.~A similar approach has been used to simulate copper and aluminum under external pressure in earlier work~\cite{lanzillo_2014}.\\

 \begin{figure}[!htb]
    \centering
	\includegraphics[width=0.48\textwidth]{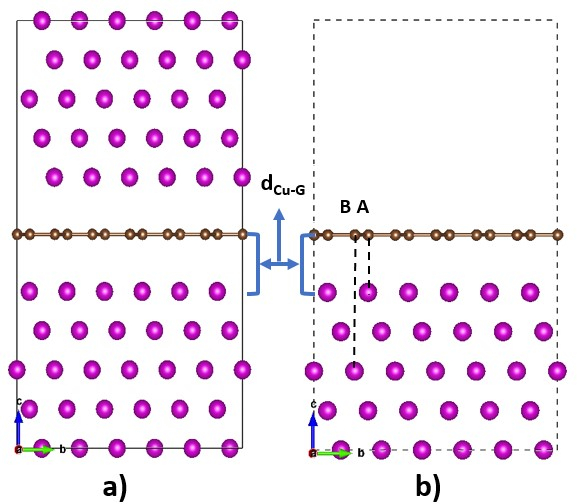}
	\label{fig_models}
    \caption{ (a)~Geometry of copper/graphene/copper (Cu/G/Cu) model with d$_{Cu-G}$ representing copper-graphene distance. b) Top-fcc configuration of graphene above the Cu 111 surface with A and B representing sub-lattice sites above the Cu atom at first and third layer respectively.~The pink colored spheres represent Cu atoms and brown colored spheres represent C atoms.}
\end{figure}

Density-functional calculations were carried out using the Vienna \textit{Ab Initio} Simulation Package \cite{K17_vasp} (VASP) code.~For atomic structure relaxations of Cu/G/Cu models, we used a plane-wave basis set with a kinetic energy cutoff of 400 eV. For static calculations, we used a larger cutoff of 520 eV.  Projected augmented wave (PAW)~\cite{K18_paw} potentials  was used to account for ion-election interactions, and the generalized gradient approximation of Perdew-Burke-Ernzerhof (PBE) \cite{K19_pbe} as the exchange-correlation functional.~All calculations were performed at 4 irreducible \textbf{k}-points, and periodic boundary conditions were implemented throughout.~To determine the partial occupancy of electrons near the Fermi level, Fermi-Dirac distribution function with a smearing temperature of 1000 K was considered.~ $\delta$ function in the expression of KGF by was approximated by a Gaussian distribution function of width 0.05 eV.~Factors like thermal broadening, \textbf{k}-point sampling, and  finite size effects discussed in references~\cite{K20,K21, K22,K23} play a role in employing the KGF. \\

To obtain representative low-energy configurations, atomic structure relaxation of the Cu/G/Cu models were performed.~The initial and final Cu-G distances after relaxation are tabulated in the first and the second column of table~\ref{table_i} respectively.~It is apparent that the extent of repulsion of Cu atoms away from the graphene is higher for models with short Cu-G distances, indicated by larger shift in Cu-G layers (third column of table ~\ref{table_i}).~For long Cu-G distances (up to $d_{Cu-G}$  = 2.61 \AA\ ), most of the Cu-Cu bonds are virtually unaffected and only few Cu-Cu bonds are formed up to $\approx$2.52 \AA, whereas for short Cu-G distances, the vertical Cu-Cu bonds are also formed at $\approx$2.49 and $\approx$2.46 \AA.\\

\begin{table}[!htb]
	\caption{\label{table_i} Variation of Cu-G distance for different Cu/G/Cu models after atomic structure relaxation and corresponding transverse component of Stress Tensor. }
	\begin{ruledtabular}
		\begin{tabular}{cccc}
			$d_{Cu-G}$\footnote{$d_{Cu-G}$ is copper-graphene distance  shown in Figure~1a) } ({\AA})  \ \ &\mbox{$d_{Cu-G}$ ({\AA}) }  &\mbox{ Shift ({\AA})} & \mbox{P$_z$\footnote{Transverse component of Stress Tensor} [GPa] }\ 
			\ \\
			 (Initial)& (Relaxed) & & \\
			\hline
			2.88&2.94&\mbox{0.06} &2.7 \\			
                2.71&2.78&\mbox{0.07} & 3.2 \\
			2.54&2.61&\mbox{0.07} &  3.1 \\
			2.37&2.46&\mbox{0.09} &   4.2\\
			2.20&2.37&\mbox{0.17} &   6.5\\
			2.02&2.29&\mbox{0.27} &10.3 \\
			1.86&2.23&\mbox{0.37} & 14.9 \\
			1.69&2.19&\mbox{0.50} &20.2 \\
		\end{tabular}
	\end{ruledtabular}
\end{table}

KGF conductivity for the Cu/G/Cu models in the direction normal to the graphene layer (i.e., along c-axis in Figure~1a) was computed.~Figure 2a displays conductivities for the models with different Cu-G distances normalized by the KGF conductivity of copper crystal computed at 300 K corresponding to $d_{Cu-Cu}$ = 2.56 {\AA}.~From Figure 2a, we see that the conductivity increases with decreasing Cu-G distance in an almost exponential manner up to $d_{Cu-G}$ = 2.37 \AA.~Beyond 2.37 \AA, we find a sharp rise in conductivity followed by a saturation below $\approx$2.23 \AA. The increase in conductivity may be attributed to increased constructive overlapping of copper and graphene orbitals with decreasing Cu-G distance.~For the short Cu-G distances with $d_{Cu-G}$ = 2.23 \AA\ and beyond, the conductivity is obtained to be $\approx$10 \% higher compared to that of copper computed at 300 K.  \\
\begin{figure}
	\includegraphics[width=0.5\textwidth]{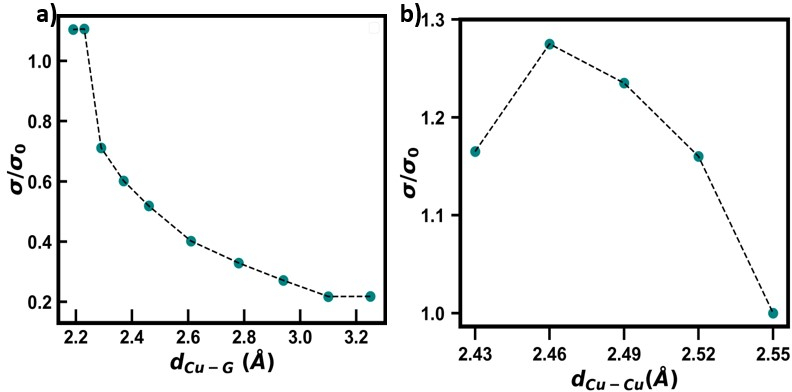}
	\caption{ a) KGF conductivities for relaxed Cu/G/Cu models with different Cu-G distance normalized by the KGF conductivity ($\sigma_{0}$) of a copper crystal computed at 300K with $d_{Cu-Cu}$ = 2.56 {\AA}. b) Relative conductivity ($\sigma / \sigma_0$) of orthorhombic Cu model computed as a function of vertical Cu-Cu bond length.}
	\label{fig:Bondlength_Cu_Cu}
\end{figure}

\begin{figure}
\centering
    \includegraphics[width=0.5\textwidth]{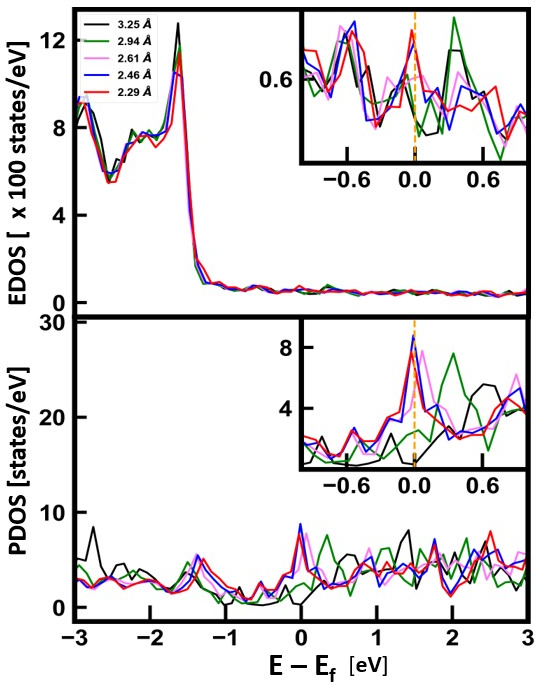}
    \caption{Electronic density of states for different Cu/G/Cu models. Top subplot corresponds to the total EDOS and the bottom subplot corresponds to the projected density of states (PDOS) onto carbon atoms.~The compression of models resulted in an enhanced EDOS near the Fermi level, see inset for details.}
    \label{fig_edos}
\end{figure}

\begin{figure*}
    \centering
    \includegraphics[width=\linewidth]{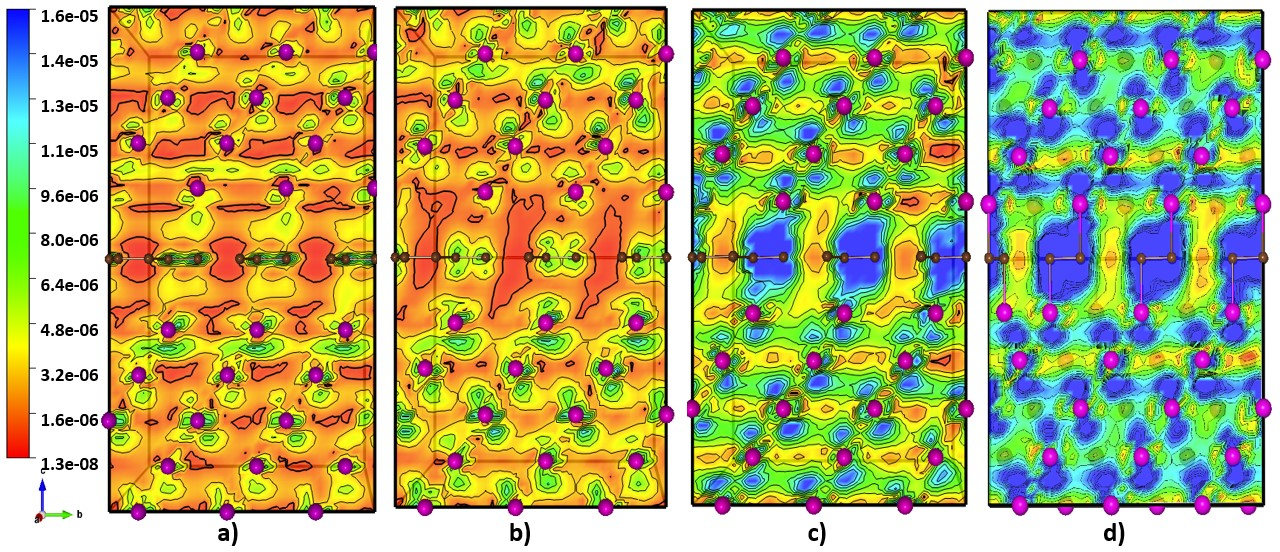}
    \caption{Space-projected conductivity plotted as a RGB colormap along a plane normal to the graphene sheet for different Cu/G/Cu models corresponding to $d_{Cu-G}$ = 3.25 {\AA}, 2.94 {\AA}, 2.29 {\AA} and 2.23 {\AA} respectively. The colorbar on the left represents the magnitude of SPC values that increases from red towards blue.~The SPC values are scaled from the minimum value for $d_{Cu-G}$ = 3.25 {\AA}. Cu and C atoms are represented by pink and brown-colored spheres respectively.}
    \label{fig_spc_selected}
\end{figure*}

To understand such an increasing trend in the conductivity with decreasing Cu-G distance, total electronic density of states (EDOS) and the projected density of states (PDOS) for selected models were computed and are shown in Figure~\ref{fig_edos}.~From Figure~\ref{fig_edos}, we find relatively higher EDOS near the Fermi level with decreasing Cu-G distance.~Such increase in the EDOS is contributed by both copper and carbon atoms of the Cu-G composites, which are more pronounced for short Cu-G distances.~According to Mott and Davis~\cite{Mott_1979}, the DC conductivity, $\sigma_{dc} \propto (N(\epsilon_{f}))^2$, where $N$ represents the density of states.~So, one expects increasing conductivity with decreasing Cu-G distance for these composites as a combined effect of enhanced EDOS near the Fermi level.~We note that it was not at all obvious that the graphene would enhance the DOS at the Fermi level, and we conjecture that this is what underlies experiments showing improved conductivity of the composites.\\

Further exploration was made by investigating a spatial description of conductivity as a function of Cu-G distance.~We computed the SPC for selected Cu/G/Cu models and are shown in Figure~\ref{fig_spc_selected}.~Figure~\ref{fig_spc_selected} displays the SPC projected on 100 plane as a RGB colormap for four such models with different Cu-G distance.~The colorbar on the left depicts the magnitude of SPC values, where red color represents low SPC values and blue color represents high SPC values.~From Figure~\ref{fig_spc_selected}, it is apparent that the contribution from both Cu and graphene to the conduction increases with decreasing Cu-G distance, and is more significant at short distances.~The SPC plots at short Cu-G distances show that the graphene directly participates in conduction and forms a bridge for conduction between Cu atoms on opposite layers.~In addition to the graphene, we also see significant contributions from Cu atoms to conduction at short Cu-G distances.~This contribution from Cu atoms to conductivity can be associated with different Cu bonding environments in the Cu-G composites.\\

 KGF conductivity for the orthorhombic Cu model for different vertical Cu-Cu distances ranging from 2.43 to 2.55 \AA were calculated.~Figure 2b shows the relative conductivity of Cu models for different Cu-Cu vertical distances.~One observes that the conductivity of Cu increases with reducing the Cu-Cu distance from $\approx$2.55 to $\approx$2.46 \AA\ after which it drops at $\approx$2.44 \AA.~We observed that for short Cu-G distances ($d_{Cu-G} \leq $ 2.23 \AA), the Cu-Cu bond-lengths are $\approx$2.49 and $\approx$2.46 Å. So, the contribution to the conductivity from the Cu atoms with nearest neighbors close to 2.46 and 2.49 \AA\ is higher compared with Cu atoms at other distances and, in agreement with SPC plots in Figure~\ref{fig_spc_selected}.\\

Saturation in  conductivity of Cu/G/Cu model for compression below $d_{Cu-G}$ = 2.23 \AA\ (refer Figure 2a) were observed.~Such a characteristic is attributed to the contribution of the Cu atoms forming different nearest-neighbor distances.~At $d_{Cu-G}$ = 2.23 \AA, the repulsion of Cu layers on both sides of graphene leads to formation of Cu-Cu bonds at $\approx$2.49 and $\approx$2.46 \AA.~With further reduction of Cu-G distance, here $d_{Cu-G}$ = 2.19 \AA, the Cu-Cu bonds are formed at $\approx$ 2.46 \AA\ and also at $\approx$2.44 \AA.~So, the additional disorder in Cu leads to slightly less contribution from the Cu atoms to the conduction (refer Figure 2b) even as the graphene still contributes.\\

In conclusion, copper-graphene composites  were modeled by considering the Cu/G/Cu interface model with varying Cu-G distance.~We showed that the DC conductivity of the Cu/G/Cu interface models increases with decreasing copper-graphene distance, and such a increase is a combined contributions from both Cu atoms and the graphene sheet.~We attribute this to an enhanced EDOS near the Fermi-level.~We also provided the spatial view of the electronic conductivity by computing the SPC for varying copper-graphene distance. The SPC calculations showed that graphene forms the bridge to electronic conduction between Cu atoms that lie on the opposite layers. We also showed an interesting characteristic of saturation of the conductivity at a short copper-graphene distance (below $d_{Cu-G}$ = 2.23 \AA), and such a characteristic is mostly attributed to slightly reduced contribution to the conduction from Cu atoms that form Cu-Cu bonds at $\approx$2.44 \AA\ even when the graphene showing enhanced contribution.
\\

This paper provides one of the first demonstrations of graphene participating in the conduction process along with the copper atoms in the composite, especially at low copper-graphene distances. An implication of this finding is that manufacturing process conditions may influence the electrical performance metal-graphene composites by engineering the interfaces to encourage graphene’s participation in the conduction processes, rather than leaving it as a scattering site. It is important to note that in addition to the interfacial distance, the arrangement of copper-carbon atoms (geometry) can also influence electrical performance of the composite; however, those effects were not explored in this work. Additional factors of importance, such as the number of graphene layers, their dimensions and the defect density of the graphene additives will be explored in future work.  
\\

\textbf{Acknowledgements:} We acknowledge Extreme Science and Engineering Discovery Environment
(XSEDE), supported by NSF Grant No. ACI-1548562 at
the Pittsburgh Supercomputing Center, for providing computational resources under allocation DMR-190008P,Department of Energy Advanced Manufacturing Office, and  Pacific
Northwest National Laboratory operated by the Battelle
Memorial Institute for the U.S. Department of Energy under
Contract No. DE-AC06-76LO1830.

\nocite{*}
\bibliography{cu-graphene}

\end{document}